\documentclass[manuscript,nonacm]{acmart}
\AtBeginDocument{%
  }

\setcopyright{cc} 
\copyrightyear{2026}
\setcctype[4.0]{by}

\begin{document}

\title{Working towards a dialectical understanding of the political ideology within technological projects}

\author{Frederick Reiber}
\email{freddyr@bu.edu}
\orcid{0000-0003-0007-5072}
\affiliation{
  \institution{Boston University}
  \city{Boston}
  \state{Massachusets}
  \country{USA}}


\begin{abstract}
\end{abstract}






\maketitle
\small This work is part of the \emph{Proceedings of CHIdeology Workshop 2026}, held on Wednesday, 15 April, in Barcelona, Spain, at CHI’26.

\section{Introduction}

Understanding the politics of technological projects is a difficult task. Technologies are rarely neutral tools; they are embedded within social struggles, normative commitments, institutional constraints, and competing visions of the future. Yet the politics of technological work often remain implicit, appearing instead as technical necessity, pragmatic compromise, or inevitable progress. In this short position paper, I develop a framework for analyzing the politics of technological projects, drawing on models from dialectical reasoning and critical social science. Rather than treating politics as something external to design or something that happens “around” technology, I argue that political commitments should be understood as internal to technological work itself. I then illustrate how this framework can be used to analyze a contemporary research agenda and conclude with reflections on what this approach contributes to the study of technological design.

\section{A Definition of Political Ideology}

In their work on developing a Marxist sociology research program, Burawoy and Wright develop an overlapping model of emancipatory social science and sociology~\cite{wright_envisioning_2020, burawoy_public_2021}. Drawing primarily from Burawoy, we can see sociology as being caught between two tensions~\cite{burawoy_public_2021}. The first is the utopian imagination, a sociology that see's the world not for what it is, but for what it could be. It is within this world that normative commitments or values---freedom, equality, and progress---can be found. Opposing the utopian imagination is the anti-utopian science. This is the realm of \textit{limitations}, where we find the constraints on human action and seek to understand how these constraints are embedded within the social institutions that we inhabit. Crucially, this is not a dystopian force that seeks to destroy utopian aims, rather, it is the analytical lens which recognizes and specifies the conditions under which our idealized values may---or may not---be realized.

As Burawoy argues, it is through the synthesis of these two tensions that we may begin to see a vision, a resolution, a path-forward. In a process akin to Hegelian dialectics, these two forces---the utopian imaginary and anti-utopian science---clash, presenting an opportunity for analysis and creating a vision of how to move forward~\cite{maybee_hegels_2020}. 

It is also within this model that I locate my definition of politics and ideology, organized around three guiding questions. The first is the question of \textbf{values}. As Burawoy argues, the first step towards any project is developing an understanding of what the world could be, a process necessitating that we draw on our normative commitments. The choice of these values---both which values we promote and in which order they are prioritized---thus constitutes the first component of my definition of ideology.

My second question concerns \textbf{constraints}, specifically the threats that one chooses to recognize. In Burawoy's formulation, anti-utopian science is recognized as the real-world limitations that impede our ability to enact our utopian vision. For example, attempts to reduce racism will put one up against the dynamics of racial capitalism that help reinforce and sustain white domination. These forces then function as anti-utopian constraints, shaping the conditions under which projects unfold. In creating my own definition of ideology, however, I shift the emphasis. Rather than treating constraints as given, I focus on which threats are recognized as legitimate or rational. This analytical move foregrounds perception, shifting our understanding from the material limitations to the ideological judgments about what one counts as "real". As David Graeber argues in \textit{The Utopia of Rules}~\cite{graeber_utopia_2015}, our understanding of reality is itself deeply contested, encoding metaphysical assumptions not actually material. The choice of which of these competing ``realities'' to inhabit---or which of these constraints to recognize as valid---then constitutes the second component of a project's ideology. 

The third question emerges from the synthesis of values and constraints, centering on \textbf{action}. As Burawoy articulates, it is through the analysis of these two forces that we can create an analysis and plan forward. For me, the central question in understanding a technological project's ideology is how this happens: how does one translate these tensions and create a theory of change. Put more explicitly, how does one create a plan to move from the world as it is to the world one seeks it to be.

\section{A Working Example}

To help illustrate how our questions and understanding of ideology play out in practice, I will now work through an example using the Mechanism Design for Social Good (MD4SG) research agenda. Originally published in 2018, Abebe and Goldner's work on Mechanism Design for Social Good argues for the use of algorithms, optimization, and mechanism design as tools for improving social outcomes~\cite{abebe_mechanism_2018}. 

\subsection{Understanding MD4SG's Values}

Understanding MD4SG's values is relatively straightforward. In the introduction to their work, Abebe and Goldner frame the agenda around the goal of improving ``access to opportunity'', illustrated through three existing research avenues.

The first concerns access to opportunities in developing communities, with the authors emphasizing the lack of reliable measures for issues such as poverty and disease, as well as the inefficiency in existing systems of trade. To address these gaps, the authors highlight projects including a gamified phone surveillance project that collects data from farmers and Kudu, an SMS-based service that facilitates transactions between farmers and buyers. These interventions---both explicit within the articles prose and implicit in their choice to focus on markets and data collection---foreground values like efficiency and allocation. The second domain focuses on online labor markets, which the authors describe as a ``rich playground'' for the study of algorithmic techniques, AI, and mechanism design. Here, the focus is on the reduction of hiring discrimination and the amount of information given to market actors. In this example, the authors focus on mitigating the social problem of discrimination, doing so through the existing labor market structures. In their final section, they discuss the allocation of housing resources, highlighting two problems. First, they discuss data analysis focused on connecting eviction to poverty, highlighting the potential for algorithms and mechanisms to improve allocation policies. The second is a direct discussion of mechanism design problems, arguing that issues with waiting lists and priority groups can be solved through the logics of optimization. Taken together, these examples reveal a consistent value orientation. Projects within MD4SG focus on improvements to systems of allocation and exchange and prioritize values like access, efficiency, and opportunity.

\subsection{Understanding MD4SG's Constraints}

MD4SG also encodes a number of constraints within its research agenda. Foremost among these is the constraint of incentive compatibility and market systems. Highlighted both in textbooks on algorithmic mechanism design and in the MD4SG paper, incentive compatibility refers to the alignment of user incentives with system designers. Put more plainly, these mechanisms assume that agents will act strategically, focusing on their own self-interest instead of the interests of the design. The designer, then, uses some sort of monetary or utility-based incentive in order to produce the desired behavior, with the underlying assumption, and thus constraint, being that human action is grounded in competition rather than cooperation.

Beyond this behavioral constraint, MD4SG also situates itself firmly within existing hegemonic economic regimes. As a research agenda, it does not challenge nor even discuss the structural conditions that generated many of the harms it seeks to address. For example, in its discussion of the developing world, the paper makes no reference to the legacies of colonialism that have produced contemporary patterns of underdevelopment, nor does it consider interventions aimed at transforming these conditions. This is also seen in the housing and labor market analysis, as the conditions of exchange---housing as marketed assets—--are treated as foundational rather than as subjects of critique or intervention. In terms of my analysis, market structures and legacies of colonization are not questions but are instead accepted as constraints within which we must devise solutions.

\subsection{Understanding MD4SG's Intervention}

The final component of my model of ideology, and the final component of MD4SG's ideology, concerns action. Specifically, what the form of \textit{proposed} interventions should take. Here, the focus is on how a project combines its values and constraints to produce a plan, and what form that plan takes---not an evaluation of how that plan plays out. Across the examples within the original paper, the articulated values---an efficiency, opportunity-driven world---collide with recognized constraints, including the limits of human nature and the persistence of systemic inequality. This tension is then resolved through technological intervention and development, functioning as the primary element of change. In the discussions of opportunity development, solutions take the form of phone SMS-based services and artificial intelligence driven systems. In the analysis of labor markets, the emphasis shifts to the design of algorithms and market mechanisms to reduce discrimination, and in the case of housing and homelessness, algorithmic intervention is presented as a central intervention. Across all three domains, social problems are rationalized with proposed solutions through technical design.

Taken together, these interventions reflect a distinct ideology. Our utopian world is one in which each individual has equal access to opportunity, systems of allocation are run efficiently through market mechanisms monitored or managed by algorithms, and progress is made through the development of new technologies, not structural transformation or collective action. 

\section{Reflections}

Understanding and defining ideology is a challenging task; articulating how technological projects embed ideology is even more challenging. While there is scholarship on modeling the politics and political economy of technology~\cite{winner_artifacts_2007, birch_artifacts_2025}, I am less aware of work that explicitly attempts to develop a model for analyzing a technology's ideology.

In developing the framework outlined here, I prioritized two commitments. First, unlike many discussions of ideology that foreground goals or values alone, I sought to treat the move from goals to values as being analytically inseparable from the constraints that form it. Ideology is often understood in terms of aspirations and actions; however, as this analysis suggests, disentangling these requires attention to the constraints recognized and dialectically encoded. Second, I aimed to develop a framework useful for studying a broad number of projects while also foregrounding technological projects, recognizing both the varied ways ideology shapes action and the focus of this workshop---disentangling the politics of HCI of technological design.

\newpage

\bibliographystyle{ACM-Reference-Format}
\bibliography{references}


\end{document}